\documentstyle[12pt,bezier]{article}
\begin{document}
\newcommand{\nc}{\newcommand}
\nc{\Vs}{v_\sigma}
\nc{\beq}{\begin{equation}}
\nc{\eeq}{\end{equation}}
\nc{\sss}{\scriptscriptstyle}
\nc{\sR}{{\sss R}}
\nc{\sL}{{\sss L}}
\nc{\sT}{{\sss T}}
\nc{\eps}{\epsilon}
\nc{\epst}{\epsilon^\sT}
\nc{\mT}{m^\sT}
\nc{\phan}{\phantom{g+\delta g}}
\nc{\ive}{i\varepsilon}
\nc{\mt}{m^{\sss T}}
\nc{\Mi}{M^{-1}}
\begin{titlepage}
\pagestyle{empty}
\baselineskip=21pt
\rightline{UMN-TH-1218-93}
\rightline{TPI-MINN-93/43-T}
\rightline{hep-ph/9309316}
\rightline{September, 1993}
\vskip .2in
\begin{center}
{\large{\bf Stability of neutrinos in the singlet majoron model}} \end{center}
\vskip .1in
\begin{center}
James M.~Cline

Kimmo Kainulainen

and

Sonia Paban

{\it School of Physics and Astronomy, University of Minnesota}

{\it Minneapolis, MN 55455, USA}

\vskip .2in

\end{center}
\vskip 1in
\centerline{ {\bf Abstract} }
\baselineskip=18pt
We show that there is no one-loop enhancement of the rate for a light
neutrino to decay into a lighter neutrino plus a majoron, contrary to a recent
claim.  Thus the light neutrinos must satisfy the cosmological bound of having
masses less than 35 eV in the singlet majoron model, or else violate the
constraint imposed by galaxy formation.  In the latter case, $\nu_\tau$ could
have a mass between 40 keV and 1 MeV, while satisfying all other cosmological
constraints.
\noindent
\end{titlepage}
\baselineskip=18pt

The singlet majoron model \cite{cmp} is one of the simplest imaginable
extensions of the standard model of particle physics.  It consists of adding
one or more right-handed neutrinos, and a complex scalar field $\sigma$ with
lepton number $-2$.  If $\sigma$ has a large vacuum expectation value $\Vs$,
one has a natural explanation for the smallness of neutrino masses through the
seesaw mechanism \cite{seesaw}.  In addition there is a massless Goldstone
boson $\chi$, the majoron, which couples primarily to neutrinos with a strength
suppressed by powers of $\Vs$.  Although the majoron would be impossible to see
directly, it could play a role in cosmological and astrophysical settings.

One of the potential uses for majorons is in evading bounds on neutrino masses.
For example a tau neutrino mass between 35 eV and 1 MeV would seriously
conflict with present estimates of the energy density of the universe since the
standard model decay $\nu_\tau\to\nu_\mu\gamma$ is too slow to deplete the
population of such neutrinos even by the present time.  The rate for decay into
a majoron rather than a photon can be many orders of magnitude faster, thus
alleviating the conflict.

In nonminimal majoron models with several scalar fields, it is certainly true
that neutrino decay rates become unsuppressed \cite{jc}.  But with only one
scalar field, it turns out that the matrix of couplings of the majoron between
the various light neutrinos is proportional to the mass matrix, at lowest order
in the small parameter $\eps\sim m_{\rm\sss Dirac}/\Vs$ \cite{sv}.  The
transition $\nu\to\nu'\chi$ only occurs at higher order in the small parameter,
which typically makes it as irrelevant as the electromagnetic decay. It is
therefore an interesting claim \cite{pal,clp} that the one-loop correction to
the amplitude for $\nu\to\nu'\chi$ is nonzero at leading order in $\eps$ and
is diminished only by typical loop suppression factors.  This claim has already
been used in the construction of at least one model \cite{bm}.

Here we wish to demonstrate the incorrectness of the claim that loop effects
enhance the neutrino decay rate.  We have done this in two different ways.
The first method is to compute the one-loop effective Lagrangian for the light
neutrinos.  Let $\nu$ denote the vector of three light neutrino states, and let
$\mu$ and $g$ denote the $3\times 3$ tree-level mass and majoron-coupling
matrices respectively.  Taking into account one-loop contributions, denoted by
$\delta\mu$ and $\delta g$, the effective Lagrangian becomes
\beq
        -\frac12\bar\nu(\mu+\delta\mu)\nu
        -\frac{i}{2}\chi\bar\nu\gamma_5(g+\delta g)\nu,
\label{e1}
\eeq
We will show that $g + \delta g$ is proportional to $\mu + \delta \mu$ at
leading order in $\eps$, just like the tree-level quantities, so that there is
no loop enhancement of the decay rate.

In the effective Lagrangian approach, only one-particle irreducible (1PI)
diagrams are included.  One must then rotate the full matrix of couplings
$g+\delta g$ to the basis where the mass matrix $\mu+\delta\mu$ is diagonal to
obtain the physical decay amplitudes.  It is also possible to calculate the
amplitude directly, including the one-particle reducible (1PR) self-energy
corrections to the tree level vertex, as in reference \cite{clp}.  An example
of such a contribution is shown in Fig.~1.  We have repeated this computation
and found discrepancies with ref.~\cite{clp}.  (We note that it is crucial to
put the internal neutrino $\nu_2$ in the 1PR diagram on the mass shell of the
external neutrino $\nu_1$ rather than at zero momentum, in order to get the
correct result.)  In this approach, the 1PR diagrams perform the function of
rotating to the one-loop mass eigenbasis.  We have verified that the 1PR and
1PI diagrams cancel each other, giving a vanishing decay amplitude at lowest
order in $\eps$.  Because the effective Lagrangian approach is simpler, we
will present our results only for that method.

Before giving the explicit calculations we observe that our result is not a
mere
accident, but can be shown by a general argument.  Consider the tree-level
Lagrangian for the singlet majoron model,
\beq
-h_{ij}{\bar L_i}HP_\sR\nu_{\sR,j}-f_{ij}\sigma\bar\nu_{\sR,i}
        P_\sR\nu_{\sR,j}
        + {\rm h.c.} - \lambda\left(|\sigma|^2-\Vs^2/2\right)^2,
\label{e2}
\eeq
where $H,L$ are the Higgs and lepton doublets, $P_\sR=(1+\gamma_5)/2$ and
$\nu_{\sR,i}$ are the isosinglet right-handed neutrinos (written in Majorana
form).  Now imagine doing the renormalizations up to any desired order in the
{\it symmetric} phase, {\it i.e.,} for negative $\Vs^2$.  The mass terms remain
zero because they are protected by the lepton number and gauge symmetries and
only the Yukawa couplings get renormalized.  The renormalized couplings should
have an analytic dependence on $\Vs^2$, so that we can compute one-loop effects
in the broken phase simply by substituting a positive value for $\Vs^2$ into
the above results.  Then the proportionality of resulting renormalized masses
and couplings is manifest just as for the tree level parameters, to {\it all}
orders in the loop expansion.  While this argument alone might suffice to prove
our point, we feel obliged to show explicitly how the correct result can be
derived in the broken phase, where the computation is more involved.

The first step is to rewrite the Lagrangian when the Higgs and $\sigma$ fields
have acquired their VEV's. There is a Dirac mass matrix $m=hv/\sqrt{2}$
connecting the left- and right-handed neutrinos, where $v=246$ GeV is the VEV
of the Higgs, and a Majorana mass matrix $M=\sqrt{2}f\Vs$ for the right-handed
neutrinos alone.  Assuming the eigenvalues of $M$ are much larger than the
entries of $m$, the full mass matrix can be partially diagonalized by
\beq
\pmatrix{\nu_\sL\cr\nu_\sR\cr} = \pmatrix{1 & \eps\cr -\epst & 1\cr}
\pmatrix{\nu\cr N\cr}; \quad \eps \equiv mM^{-1},
\label{e3}
\eeq
so that $\nu$ and $N$ respectively have the light and heavy mass matrices
$\mu=-m M^{-1}\mT$ and $M$, to leading order in $\eps$.
We also expand
\beq
\sigma = \frac{1}{\sqrt{2}}\left(\Vs + \rho + i\chi\right).
\label{e4}
\eeq
In constrast to the massless majoron $\chi$, $\rho$ has a mass given by
$m^2_\rho = 2\lambda\Vs^2$.  The interaction Lagrangian is then
\beq
{\cal L}_i = -\frac{1}{2\Vs}\pmatrix{\bar \nu &\bar N\cr}
        \pmatrix{\eps{M}\epst & -\eps{M}\cr
                -{M}\epst & {M} \cr}
        (\rho+i\gamma_5\chi) \pmatrix{\nu\cr N\cr}
        -\frac{\lambda}{4}(2\Vs\rho+\rho^2+\chi^2)^2,
\label{e5}
\eeq
from which the Feynman rules are easily deduced.  The above-emphasized
proportionality of the $\nu\nu\chi$ coupling matrix to the light mass matrix
$\mu$ is manifest:
\beq
        g = {\eps M\epst\over\Vs} = -\frac{\mu}{\Vs}.
\label{e6}
\eeq

The next step is to examine the general structure of the one-loop shifts in the
mass and coupling matrices.  In the complete space of the light and heavy
neutrinos, $(\nu, N)$, the masses have the form
\beq
        {\rm masses\ } = \frac12\pmatrix{\mu+\delta\mu & \delta m \cr
                                  \delta\mT & M + \delta M \cr}.
\label{e7}
\eeq
which necessitates rediagonalizing the light and heavy states using the
transformation
\beq
        O = \pmatrix{1&\delta\eps\cr -\delta\epst & 1};
        \quad \delta\eps = \delta m M^{-1}.
\label{e8}
\eeq
Here and below we have kept only the terms which are of leading order in the
combined $\eps$ and loop expansion.  (For example, keeping $\delta M$ in
the definition of $\delta\eps$ would be consistent only if we were doing a
two-loop calculation.)  After the rotation (\ref{e8}), $\mu+\delta\mu$ gets a
further shift $-\delta m^\sT M^{-1}\delta m$ which is equivalent to a two-loop
effect and can be ignored.  But there is a nonnegligible shift in the matrix of
couplings, which upon rediagonalization undergoes
\beq
        {\rm couplings\ } = \frac12\pmatrix{g+\delta g & g'+\delta g'\cr
                                g'^\sT+\delta g'^\sT & G+\delta G\cr}\to
        \frac12\pmatrix{g+\delta g - g'\delta\epst - \delta\eps g'^\sT & *\cr
                        \phan *\phan & \phan *\phan \cr},
\label{e9}
\eeq
where $*$ indicates the terms we are not interested in.  Thus what we called
$\delta g$ in eq.~(\ref{e1}) is actually given by the naive one-loop shift plus
terms due to heavy-light mixing at one loop.  {\it We will henceforth use
$\delta g$ to denote the `naive' contribution.} From this analysis we see that
it is necessary only to compute the quantities $\delta g$, $\delta m$, and
$\delta \mu$.  These are given, respectively, by the Feynman diagrams shown in
figures 2, 3 and 4.

First consider $\delta g$, figs.~2a-b.  Although there are additional
digrams not shown, with $\nu$ rather than $N$ running in the loop, they
are of higher order in $\eps$.  Adding the contributions of fig.~2 evaluated at
vanishing external momenta, we obtain the matrix equation
\beq
\delta g = m^2_\rho\Vs^{-3}\eps I(M,m_\rho )\epst,
\label{e10a}
\eeq
with
\beq
I(x,y ) = {1 \over 16\pi^2}{x^3\over x^2-y^2}
\ln \frac{x^2}{y^2}.
\label{e10b}
\eeq
The factor of $m^2_\rho$ in the numerator here and below always comes from the
difference between the propagators of $\chi$ and $\rho$, or from the fact that
the $\chi\chi\rho$ coupling is proportional to $m^2_\rho$.  Counting powers of
$\Vs$ (remember that $m_\rho\sim M\sim\Vs$), we see that $\delta g$ is of order
$\eps^2\sim (m/M)^2$, and is not proportional to the tree level mass matrix
$\mu$.  If this were the end of the story, $\delta g$ would cause fast decays
among the light neutrinos.  But we must also consider the mass shifts.

The Dirac mass shift $\delta m$ is given by fig.~3.  A straightforward
calculation shows that
\beq
\delta m = - m^2_\rho\Vs^{-2} I(M,m_\rho ) \epst.
\label{e11}
\eeq
Using $\delta\eps=\delta m M^{-1}$ from eq.~(\ref{e8}) and $g'=-\eps M/\Vs$
from eq.~(\ref{e5}), we find that
\beq
\delta g_{\rm\sss TOT} = \delta g -g'\delta\epst-\delta\eps g'^\sT = -\delta g,
\label{e12}
\eeq
so the net effect of rotating away the induced heavy-light mixing is to reverse
the sign of the naive shift in the couplings.

The final singlet Higgs contribution is the direct shift in the Majorana mass
matrix of the light neutrinos, fig.~4, which we find to be given by $\delta\mu
= \Vs\delta g$. Therefore the relation between the coupling and mass matrices
at one loop and leading order in $\eps$ is
\beq
g+\delta g_{\rm\sss TOT} = - {\mu+\delta\mu\over\Vs},
\label{e13}
\eeq
that is, they are exactly proportional to each other.  Thus they are
simultaneously diagonalized and there is no loop-enhancement of decays.

{}From the general argument given above, it should be clear that our results do
not depend on what kind of interactions are involved, so long as they do not
explicitly break the global lepton number symmetry.  Thus the one-loop effects
involving Higgs and weak gauge bosons must also preserve the proportionality
between masses and couplings.  We have also carried out this calculation.

\nc{\msz}{M^2_{\sss Z}}
\nc{\msh}{M^2_{\sss H}}

The relevant diagrams are shown in figs.~5 and 6 for $\delta g$ and $\delta
\mu$ respectively.  In contrast to the previous case, simple power counting
(taking into account necessary insertions of neutrino masses) shows that the
electroweak contribution to $\delta m$ is $O(\eps^3)$ and so can be neglected.
The $Z$-boson interactions may be written as
\beq
{-g\over 4\cos\theta_{\sss W}}\pmatrix{\bar\nu & \bar N\cr}
\pmatrix{1 & \eps \cr \epst & \epst\eps \cr} \gamma_\mu\gamma_5
\pmatrix{\nu\cr N\cr} Z^\mu,
\label{e14}
\eeq
using the fact that the vector current vanishes for Majorana fields.  In
evaluating the diagrams it proves convenient to work in $R_\xi$ gauge with $\xi
= \msh/\msz$, for then the contributions from the real and imaginary part of
the Higgs field exactly cancel each other, and we are left with only the gauge
boson diagrams.  We obtain
\begin{eqnarray}
\delta\mu  = -{g^2\over 4\cos^2\theta_{\sss W}}
\eps \left( 3I(M,M_{\sss Z}) + (\msh/\msz) I(M,M_{\sss H}) \right) \epst,
\label{e15a}
\end{eqnarray}
where
$I$ is given in eq.~(\ref{e10b}); notice that the relative contributions of the
Higgs boson and the three polarizations of the Z are evident. We also find that
\beq
\delta g_{\rm\sss TOT} = \delta g = - {\delta \mu\over \Vs},
\label{e16}
\eeq
explicitly showing that the weak interactions also preserve the proportionality
between masses and couplings that forbids transitions between light neutrinos
at $O(\eps^2)$ in the amplitude.

(While the contribution of fig.~7 might at first appear to be
relevant, it is $O(\eps^4)$.  The subgraph containing $Z_\mu$-$\chi$ mixing is
of order $\eps^2 q_\mu$, where $q_\mu$ is the majoron momentum.  Integrating by
parts the effective operator $\bar\nu\gamma_\mu\nu\partial_\mu\chi$ and using
the equations of motion gives another factor of the light neutrino mass,
$\mu\sim\eps^2$.)

Finally, we examine the tree level decay rate more explicitly than appears to
have been done elsewhere.  With a little effort, one can go to next order in
$\eps$ in the diagonalization of the neutrino mass matrix, to find that the
off-diagonal block of the heavy-light rotation matrix, $\eps$ in
eq.~(\ref{e3}), should be supplemented by the term $-m\Mi(\mt m\Mi + \frac12
\Mi\mt m)\Mi$.  This is enough to determine the masses and majoron couplings
for the light neutrinos to order $\eps^4$,
\beq
    \mu = \mu_0 - \frac{1}{2}\left\{\mu_0,\eps\epst\right\},\qquad
    g = -\frac{\mu}{\Vs} + \left\{\frac{\mu}{\Vs},\eps\epst\right\}
\label{e19}
\eeq
where $\mu_0 = -\eps M\epst$. To more easily explore the consequences of this
result, let us assume there is mixing only among the last two generations, with
eigenvalues $m_{\nu_i}$ and $M_i$ for the light and heavy masses.  In the mass
basis, one can show that the off-diagonal coupling in eq.~(\ref{e19}) is given
by
\beq
        g_{23} = \frac{1}{2}\sin 2\alpha\left(M_2^{-1}-M_3^{-1}\right)
        \left(m_{\nu_2}m_{\nu_3}\right)^{1/2}(m_{\nu_2}+m_{\nu_3})/\Vs,
\label{e20}
\eeq
where $\alpha$ is the rotation angle for diagonalizing the matrix $M^{-1/2}\mt
mM^{-1/2}$.  (Note that this vanishes if the heavy neutrinos are degenerate,
as expected, since if $M$ is proportional to the unit matrix, the Dirac
masses can be diagonalized simultaneously with it, in which case there is no
mixing or decay.) For the purpose of obtaining cosmological upper bounds on
$\Vs$, we conservatively assume that the angle is large and the eigenvalues
$M_i$ have a large splitting, so that $g_{23}\sim
m_{\nu_2}^{1/2}m_{\nu_3}^{3/2}/\Vs^2$, leading to a decay rate of
\beq
        {1\over\tau} \sim {m_{\nu_2} m_{\nu_3}^4 \over 16\pi\Vs^4}
\label{e21}
\eeq
for the process $\nu_3\to\nu_2\chi$.

If $m_{\nu_3}$ exceeds the cosmological upper bound of 35 eV for stable
neutrinos \cite{blud}, $\nu_3$ must decay fast enough so that its
relativistic decay products have time for their energy to be redshifted away.
{}From eq.~(\ref{e21}) it is clear that we want $m_{\nu_2}$ to be as large as
possible.  Demanding that the present energy density of $\nu_2$ particles plus
majorons from the decaying $\nu_3$ not exceed the closure value, we obtain a
bound involving the decay temperature $T_{\sss D}$,
\beq
	m_{\nu_2} + \left(m_{\nu_2}^2 + p^2\right)^{1/2} + p < 35 {\rm\ eV\ };
	\qquad p = \frac{1}{2}{m_{\nu_3}T_0\over T_{\sss D}},
\label{e22}
\eeq
where $p$ is the redshifted momentum of the majoron.
With the aid of eq.~(\ref{e21}) and the time-temperature relationship
for a universe dominated by nonrelativistic $\nu_3$'s, we get
\beq
	\Vs < 500{\rm\ GeV\ }\left({m_{\nu_3}\over 1{\rm\ MeV}}\right)^{1/2},
\label{e23}
\eeq
assuming $m_{\nu_2}$ has the optimal value ($\sim 9$ eV) for giving the least
restrictive bound.

If $m_{\nu_3}$ was smaller than 240 keV, this would imply that $\Vs$ is below
the weak scale of 246 GeV, a somewhat small value from the perspective of the
seesaw mechanism for the neutrino masses.  Nevertheless with $\Vs=100$ GeV, the
Dirac neutrino masses corresponding to $m_{\nu_1}\sim 1$ eV, $m_{\nu_2}\sim 10$
eV, $m_{\nu_3} \sim 40$ keV are $0.3$ MeV, $1$ MeV and $60$ MeV, respectively,
comparable to the smallest lepton and quark masses.  If $m_{\nu_3}$ saturated
the nucleosynthesis bound of 500 keV \cite{kolb} then $\Vs$ could be as large
as 350 GeV and the neutrino Dirac masses could span a slightly greater range,
up to 0.4 GeV for the third generation.

If the majoron has a gravitationally induced mass \cite{akh}, whose natural
value is around 1 keV, the above results still hold, except for the fact that
now the final density of $\nu_2$ will be four times its thermal value (due to
the decay products of $\chi\to\nu_\mu\nu_\mu$), so that its mass should be
somewhat smaller.  The ensuing bound on $\Vs$ decreases only slightly.

Although considerations of the cosmological energy density leave room for the
relevance of light neutrino decays in the singlet majoron model, the growth of
large scale structure leads to a different conclusion.  The decay products of
a massive tau neutrino could cause a second radiation-dominated era, during
which density perturbations grow only logarithmically with time.  The recent
measurements of COBE \cite{cobe} indicate that any additional period of
radiation domination would conflict with the necessity for primordial
perturbations to have grown into galaxies by now.  Therefore $\nu_\tau$ should
never have matter-dominated the energy density of the universe \cite{steig},
hence the decay temperature of $\nu_\tau$ must not be much less than its mass
(more precisely, $m_{\nu_3}/T_{\sss D} < 14$; see ref.~\cite{osv}), leading to
the constraint
\beq
   \Vs < 20 {\rm\ GeV\ }\left({m_{\nu_3}\over 1 {\rm\ MeV}}\right)^{1/2}
    \left({m_{\nu_2}\over 35{\rm\ eV}}\right)^{1/4}.
\label{e24}
\eeq
For any allowed value of $m_{\nu_3}$, this is too small a value of $\Vs$ to
give plausible seesaw masses for the light neutrinos.

In summary, we have given both a simple general argument valid to all orders
in perturbation theory, and detailed calculations at one loop in the broken
phase, to show that decay amplitudes between light neutrinos always vanish at
order $\eps^2$ in the simplest majoron model, where $\eps$ is the ratio between
the scales of Dirac and Majorana masses of the neutrinos. Thus there is no
enhancement of the  amplitude $\nu \to \nu '\chi$ due to loop
corrections in the simplest majoron model.  We have also shown that the tree
level decays are too slow to be cosmologically relevant, but only if the galaxy
formation constraint is considered.
\vskip 0.4in
\noindent{ {\bf Acknowledgements} } \\
\noindent{ This work was supported in part by DOE grant DE-AC02-83ER-40105.}

\newpage
\begin{center} {\large Figure Captions} \end{center}
\begin{description}
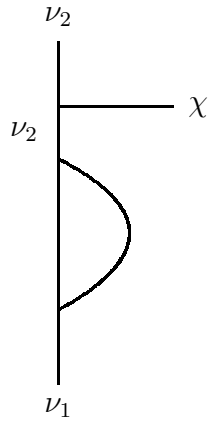
\item[Fig.~1] One-particle reducible contribution to the decay amplitude.
\item[Fig.~2a,b] Singlet Higgs contributions to $\delta g$.
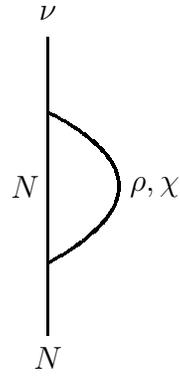
\item[Fig.~3] Singlet Higgs contribution to $\delta m$.
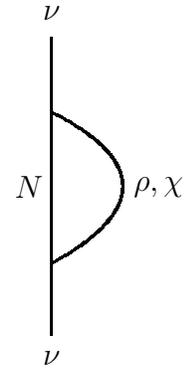
\item[Fig.~4] Singlet Higgs contribution to $\delta \mu$.
\item[Fig.~5a,b] Electroweak contributions to $\delta g$,  using all
four combinations of $\nu$ and $N$ in 5b.
\item[Fig.~6a,b] Electroweak contributions to $\delta \mu$.
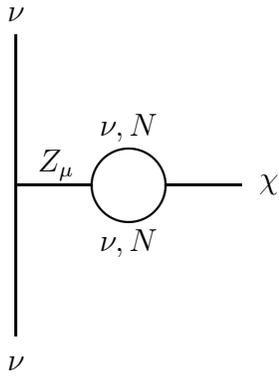
\item[Fig.~7] A subleading contribution to $\delta g$.
\end{description}

\unitlength=1.00mm
\thicklines
\begin{picture}(99.33,18.30)(0,20)
\put(80.00,15.00){\line(0,-1){45.61}}
\bezier{168}(80.00,-0.67)(98.67,-10.28)(80.00,-20.67)
\put(80.03,18.30){\makebox(0,0)[cc]{$\nu_2$}}
\put(80.03,-33.69){\makebox(0,0)[cc]{$\nu_1$}}
\put(80.00,-43.41){\makebox(0,0)[cc]{Fig.~1}}
\put(80.00,6.33){\line(1,0){15.33}}
\put(97.33,6.33){\makebox(0,0)[lc]{$\chi$}}
\put(77.33,3.00){\makebox(0,0)[rc]{$\nu_2$}}
\end{picture}

\newpage
\unitlength=1.00mm
\thicklines
\begin{picture}(148.67,133.30)(-10,-20)
\put(10.00,130.00){\line(0,-1){39.61}}
\bezier{168}(10.00,120.00)(28.67,110.39)(10.00,100.00)
\put(50.00,130.00){\line(0,-1){39.61}}
\bezier{168}(50.00,120.00)(68.67,110.39)(50.00,100.00)
\put(90.00,130.00){\line(0,-1){39.61}}
\bezier{168}(90.00,120.00)(108.67,110.39)(90.00,100.00)
\put(10.00,110.00){\line(-1,0){9.97}}
\put(-3.65,110.14){\makebox(0,0)[cc]{$\chi$}}
\put(6.79,115.02){\makebox(0,0)[cc]{$N$}}
\put(6.79,105.58){\makebox(0,0)[cc]{$N$}}
\put(10.05,133.29){\makebox(0,0)[cc]{$\nu$}}
\put(10.05,87.00){\makebox(0,0)[cc]{$\nu$}}
\put(21.22,110.14){\makebox(0,0)[lc]{$\rho,\chi$}}
\put(59.29,110.14){\line(1,0){10.66}}
\put(73.12,110.14){\makebox(0,0)[cc]{$\chi$}}
\put(46.61,110.14){\makebox(0,0)[cc]{$N$}}
\put(50.03,133.29){\makebox(0,0)[cc]{$\nu$}}
\put(50.03,87.00){\makebox(0,0)[cc]{$\nu$}}
\put(57.77,118.67){\makebox(0,0)[cc]{$\rho$}}
\put(57.77,101.92){\makebox(0,0)[cc]{$\chi$}}
\put(10.00,77.59){\makebox(0,0)[cc]{Fig.~2a}}
\put(50.00,77.59){\makebox(0,0)[cc]{Fig.~2b}}
\put(90.03,133.30){\makebox(0,0)[cc]{$\nu$}}
\put(90.03,87.31){\makebox(0,0)[cc]{$N$}}
\put(86.99,110.15){\makebox(0,0)[cc]{$N$}}
\put(100.94,110.15){\makebox(0,0)[lc]{$\rho,\chi$}}
\put(130.00,130.00){\line(0,-1){39.61}}
\bezier{168}(130.00,120.00)(148.67,110.39)(130.00,100.00)
\put(130.03,133.30){\makebox(0,0)[cc]{$\nu$}}
\put(130.03,87.31){\makebox(0,0)[cc]{$\nu$}}
\put(126.99,110.15){\makebox(0,0)[cc]{$N$}}
\put(140.94,110.15){\makebox(0,0)[lc]{$\rho,\chi$}}
\put(90.00,77.59){\makebox(0,0)[cc]{Fig.~3}}
\put(130.00,77.59){\makebox(0,0)[cc]{Fig.~4}}
\put(10.00,2.59){\makebox(0,0)[cc]{Fig.~5a}}
\put(50.00,2.59){\makebox(0,0)[cc]{Fig.~5b}}
\put(10.00,55.00){\line(0,-1){39.61}}
\bezier{168}(10.00,45.00)(28.67,35.39)(10.00,25.00)
\put(10.00,35.00){\line(-1,0){9.97}}
\put(6.79,40.02){\makebox(0,0)[cc]{$N$}}
\put(6.79,30.58){\makebox(0,0)[cc]{$N$}}
\put(10.05,58.29){\makebox(0,0)[cc]{$\nu$}}
\put(10.05,11.08){\makebox(0,0)[cc]{$\nu$}}
\put(21.22,35.14){\makebox(0,0)[lc]{$H$}}
\put(50.00,55.00){\line(0,-1){39.61}}
\bezier{168}(50.00,45.00)(68.67,35.39)(50.00,25.00)
\put(50.00,35.00){\line(-1,0){11.97}}
\put(36.35,35.14){\makebox(0,0)[cc]{$\chi$}}
\put(48.89,40.02){\makebox(0,0)[rc]{$\nu,N$}}
\put(48.89,30.58){\makebox(0,0)[rc]{$\nu,N$}}
\put(50.05,58.29){\makebox(0,0)[cc]{$\nu$}}
\put(50.05,11.08){\makebox(0,0)[cc]{$\nu$}}
\put(61.22,35.14){\makebox(0,0)[lc]{$Z_\mu$}}
\put(90.00,55.00){\line(0,-1){39.61}}
\bezier{168}(90.00,45.00)(108.67,35.39)(90.00,25.00)
\put(90.03,58.30){\makebox(0,0)[cc]{$\nu$}}
\put(90.03,12.31){\makebox(0,0)[cc]{$\nu$}}
\put(86.99,35.15){\makebox(0,0)[cc]{$N$}}
\put(100.94,35.15){\makebox(0,0)[lc]{$H$}}
\put(130.00,55.00){\line(0,-1){39.61}}
\bezier{168}(130.00,45.00)(148.67,35.39)(130.00,25.00)
\put(130.03,58.30){\makebox(0,0)[cc]{$\nu$}}
\put(130.03,12.31){\makebox(0,0)[cc]{$\nu$}}
\put(128.99,35.15){\makebox(0,0)[rc]{$\nu,N$}}
\put(140.94,35.15){\makebox(0,0)[lc]{$Z_\mu$}}
\put(90.00,2.59){\makebox(0,0)[cc]{Fig.~6a}}
\put(130.00,2.59){\makebox(0,0)[cc]{Fig.~6b}}
\put(10.00,-20.00){\line(0,-1){40.00}}
\put(10.00,-40.00){\line(1,0){10.00}}
\put(25.00,-40.00){\circle{10.00}}
\put(30.00,-40.00){\line(1,0){10.00}}
\put(10.00,-17.33){\makebox(0,0)[cc]{$\nu$}}
\put(10.00,-63.67){\makebox(0,0)[cc]{$\nu$}}
\put(15.33,-37.33){\makebox(0,0)[cc]{$Z_\mu$}}
\put(25.00,-32.33){\makebox(0,0)[cc]{$\nu,N$}}
\put(25.00,-47.67){\makebox(0,0)[cc]{$\nu,N$}}
\put(-4.00,34.67){\makebox(0,0)[lc]{$\chi$}}
\put(42.34,-40.00){\makebox(0,0)[lc]{$\chi$}}
\put(10.00,-70.00){\makebox(0,0)[cc]{Fig.~7}}
\end{picture}

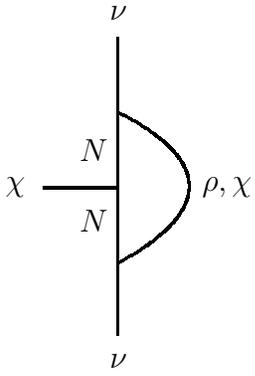
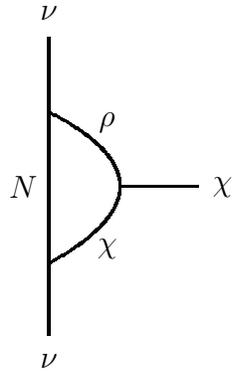
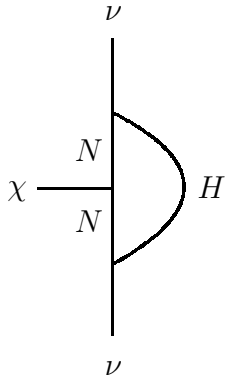
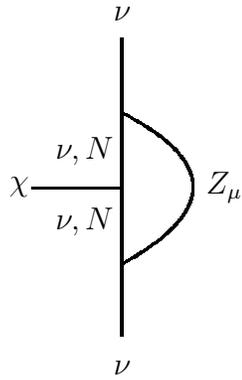
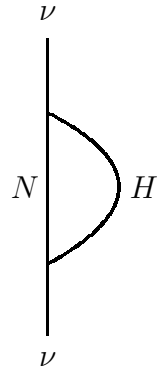
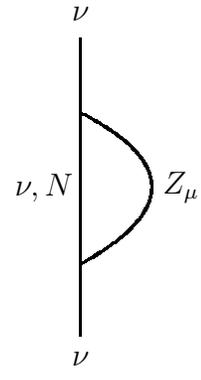
\end{document}